\documentclass[runningheads]{llncs}
\usepackage[T1]{fontenc}
\usepackage{graphicx}

\usepackage[T1]{fontenc}
\usepackage{graphicx}
\usepackage{booktabs}
\usepackage[misc]{ifsym}
\usepackage{amsmath}
\usepackage{algorithm}
\usepackage[switch]{lineno}
\usepackage{multirow}
\usepackage{wrapfig}
\usepackage{subcaption}
\usepackage{array}
\usepackage{algpseudocode}
\usepackage{amssymb}

\newcommand{\corr}{(\Letter)}
\begin{document}
\title{Action-Attentive Deep Reinforcement Learning for Autonomous Alignment of Beamlines}

\titlerunning{Deep Reinforcement Learning for Autonomous Alignment of Beamlines}

\author{Siyu Wang \and
Shengran Dai  \and Jianhui Jiang \and Shuang Wu \and \\ Yufei Peng \and Junbin Zhang \corr}

\authorrunning{S. Wang et al.}

\institute{Gusu Laboratory of Materials, Suzhou, China\\
\email{\{wangsiyu2022,zhangjunbin2021\}@gusulab.ac.cn}\\}

\maketitle              
\begin{abstract}
Synchrotron radiation sources play a crucial role in fields such as materials science, biology, and chemistry. The beamline, a key subsystem of the synchrotron, modulates and directs the radiation to the sample for analysis. However, the alignment of beamlines is a complex and time-consuming process, primarily carried out manually by experienced engineers. Even minor misalignments in optical components can significantly affect the beam's properties, leading to suboptimal experimental outcomes. Current automated methods, such as bayesian optimization (BO) and reinforcement learning (RL),
 although these methods enhance performance, limitations remain. The relationship between the current and target beam properties, crucial for determining the adjustment, is not fully considered. Additionally, the physical characteristics of optical elements are overlooked, such as the need to adjust specific devices to control the output beam's spot size or position. 
This paper \footnote{Our code is available at https://github.com/sygogo/alignment\_beamlines\_rl} addresses the alignment of beamlines by modeling it as a Markov Decision Process (MDP) and training an intelligent agent using RL. The agent calculates adjustment values based on the current and target beam states, executes actions, and iterates until optimal parameters are achieved. A policy network with action attention is designed to improve decision-making by considering both state differences and the impact of optical components. Experiments on two simulated beamlines demonstrate that our algorithm outperforms existing methods, with ablation studies highlighting the effectiveness of the action attention-based policy network. 

\end{abstract}
\section{Introduction}

\begin{figure}
\includegraphics[width=\textwidth]{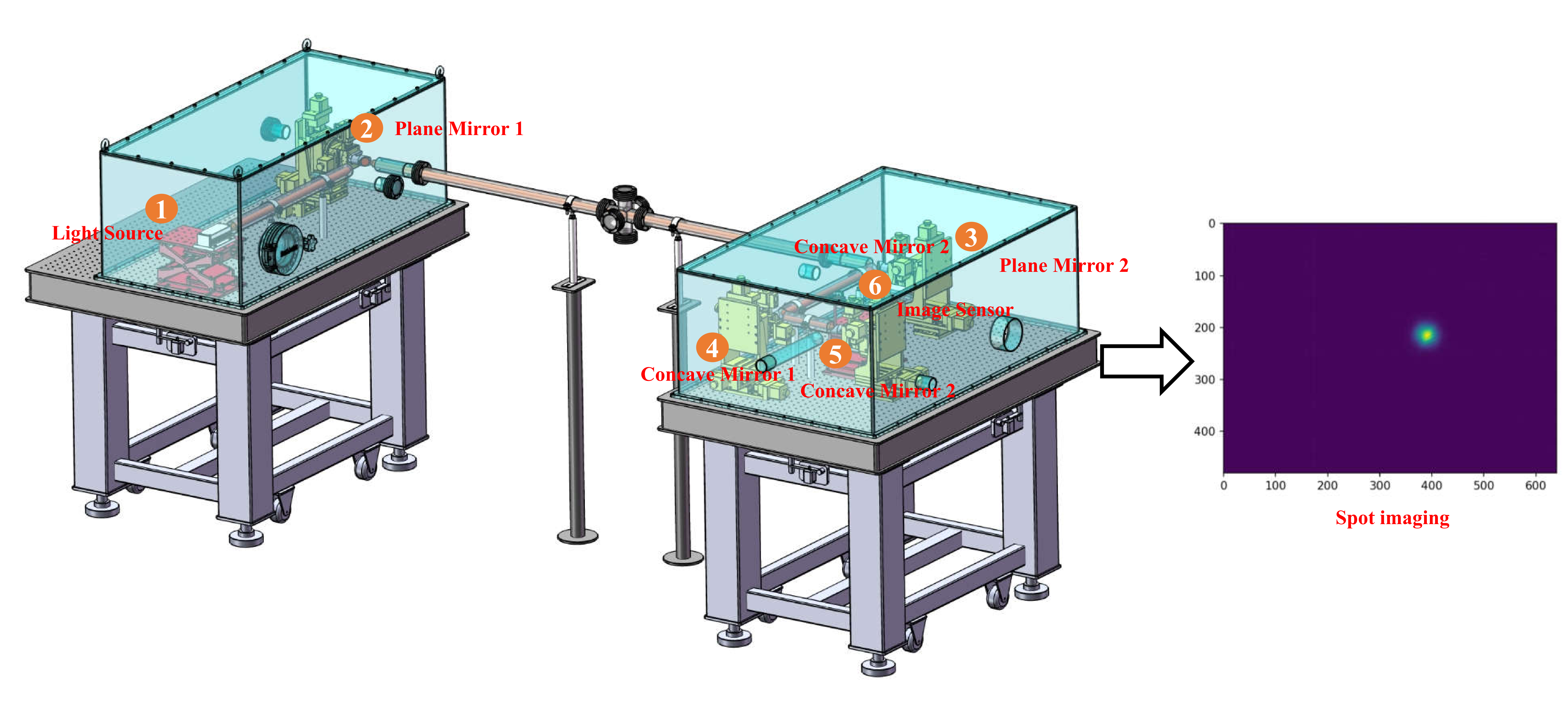}
\caption{A simple beamline. It includes 1 light source, 4 optical devices, and 1 detector. The optical device is used to transform the light emitted by the light source and finally present it to the detector.} \label{fig1}
\end{figure}

A synchrotron radiation source is an extremely bright light source that can produce a wide spectrum of electromagnetic radiation, including photons from infrared to X-rays \cite{garcia2016synchrotron}. At present, it is mainly used for research in the fields of materials science, biology, chemistry, etc., for experiments such as fine structure analysis, imaging, spectroscopy and material properties testing. The beamline is a key subsystem of a synchrotron radiation source. It characteristically modulates the light source and transmits it to the experimental states for research. The beamlines function is similar to series-connected electrical circuits, where any malfunctioning component can prevent the synchrotron beam from reaching the sample. Even minor changes in the angle or position of an optical element can have significant effects \cite{karaca2024optimization}. The beamline usually includes reflectors, monochromators, focusing mirrors and detectors \cite{karaca2024optimization}. By controlling and adjusting these optical devices, the beam of the synchrotron radiation source is adjusted (including the intensity, energy, direction, and size of the light) to ensure that the beam can interact with the sample in the best way and obtain high-quality experimental data. Currently, the adjustment of the beamline mainly relies on experienced engineers who control equipment prudently, and it is a very time-consuming and labor-intensive process.

With the development of artificial intelligence technology, in recent years, some studies have used combinatorial optimization algorithms such as bayesian optimization algorithms \cite{kaiser2023learning}, genetic algorithms \cite{morris2024general,hwang2022beam,morris2022fly} and reinforcement learning \cite{chen2023trend} to achieve automatic adjustment of optical elements in beamlines, thereby helping experimenters quickly obtain an ideal experimental environment. These studies regard the autonomous alignment of beamlines as a combinatorial optimization problem, adjusting optical elements through different algorithms to output the beam desired by the experimenter. Although these methods significantly enhance performance, certain limitations remain. (1) They do not fully account for the relationship between the current output beam's properties (current state) and the desired beam's properties (target state), which are critical for subsequent adjustments. When the current state deviates significantly from the target state, a large adjustment is required; otherwise, a smaller adjustment is sufficient. (2) They overlook the physical characteristics of different optical elements. As illustrated in Figure 1, to modify the spot size of the output beam, one should primarily adjust the position and angle of devices 4 and 5. Conversely, to adjust the position of the output beam, the focus should be on altering the position and angle of optical devices 2 and 3.

To handle the above-mentioned issues, this paper first regards the autonomous alignment of beamlines as a Markov Decision Process (MDP) \cite{puterman2014markov} and trains an intelligent agent through reinforcement learning. The intelligent agent combines the user's expected target state and the current state to calculate the next action (adjustment value), executes the action to obtain a new state, and then repeats the whole process until the optimal parameters are found. To enable the agent to perceive the difference between the target and the current state and the impact of optical components on the light beam when making decisions and generating more reasonable actions, this paper designs a policy network based on action attention to generate the actions of the intelligent agent. Finally, to verify the effectiveness of the algorithm, we built two small beamlines and simulated the input light source through a laser transmitter.
Experiments in two simulated systems show that our algorithm can achieve better performance than other methods. At the same time, ablation experiments prove that the policy network based on action attention can better generate the next action value of the agent in this task. 

The contributions of this paper include: (1) The autonomous alignment of beamlines is regarded as a Markov Decision Process (MDP), and the agent is trained through reinforcement learning. (2) A policy model based on action attention is designed, which enables the agent to adjust different optical devices differently according to the target output. (3) Two simulated small beamlines are constructed, and the effectiveness of our method is verified by experiments in the simulated beamlines.

\section{Related Works}

Currently, beamline alignment relies heavily on skilled engineers to manually control the equipment, the process is both time-consuming and labor-intensive. With advances in artificial intelligence, combinatorial optimization methods such as bayesian optimization and genetic algorithms are increasingly applied to automate the adjustment of optical components in beamlines, allowing researchers to achieve optimal experimental conditions more efficiently.

\subsection{Optimization Algorithm in Beamlines Alignment}

\cite{morris2024general} developed a streamlined software framework for beamline alignment, which was tested across four distinct optimization problems relevant to experiments at the X-ray beamlines of the National Synchrotron Light Source II and the Advanced Light Source, as well as an electron beam at the Accelerator Test Facility. They also conducted benchmarking using a simulated digital twin. The study discusses novel applications of this framework and explores the potential for a unified approach to beamlines alignment across various synchrotron facilities. \cite{morris2022fly} developed an online learning model for autonomous optimization of optical parameters using data collected from the Tender Energy X-ray Absorption Spectroscopy (TES) beamline at the National Synchrotron Light Source-II (NSLS-II). \cite{zhang2023multi} introduced a novel optimization method based on a multi-objective genetic algorithm, and they attempted to optimize a beamline with multiple objectives.
\cite{karaca2024optimization} investigated the performance of different evolutionary algorithms on the beamline calibration task. 

In recent years, deep reinforcement learning has achieved good results in combinatorial optimization \cite{mazyavkina2021reinforcement}.
\cite{hwang2022beam} presented their initial efforts toward applying machine learning (ML) for the automatic control of the beam exiting the front end (FE). They develop and test a prior-mean-assisted bayesian optimization (pmBO) method, where the prior model is trained using historical or archived data. \cite{kaiser2023learning} conducted a comparative study using a routine task in a real particle accelerator as an example, demonstrating that reinforcement learning-based optimization (RLO) generally outperforms bayesian optimization (BO), although it is not always the optimal choice. Based on the results of this study, they provided a clear set of criteria to guide the selection of the appropriate algorithm for specific tuning tasks. Lasted study \cite{chen2023trend} proposed a trend-based soft actor-critic(TBSAC) beam control method with strong robustness, allowing the agents to be trained in a simulated environment and applied to the real accelerator directly with zero-shot.

\subsection{Optimization Algorithm in Synchrotron Radiation Source}

In addition to being widely used in beamlines, optimization algorithms also have many application scenarios in other components of synchrotron radiation sources. For example, \cite{meier2012orbit} employed an actor-critic framework to correct the trajectory of a storage ring in a simulated environment. Furthermore, reinforcement learning \cite{boltz2020feedback} is also implemented to stabilize the operation of THz CSR (Terahertz Coherent Synchrotron Radiation) in synchrotron light sources, overcoming instability limitations caused by bunch self-interaction. \cite{ruichun2021application} and \cite{schirmer2019orbit} trained controllers based on historical Beam Position Monitor data to realize online orbit correction in synchrotron light sources.

\section{Problem Formulation}

The autonomous alignment of beamlines can be conceptualized as a MDP \cite{puterman2014markov}, wherein the agent continuously interacts with its environment. Specifically, the agent assesses the current state of the environment and generates a control signal based on this state. In response, the environment returns a new state to the agent along with reward information. Subsequently, the agent updates its policy according to the reward received from the environment. Thus, the primary objective of reinforcement learning is to derive the optimal policy that maximizes cumulative rewards. The following is a formal definition of reinforcement learning.

\textbf{Agent:} The agent perceives the state of the external environment and the rewards fed back, and learns and makes decisions. The decision-making function of the agent refers to taking different actions according to the state of the external environment, and the learning function refers to adjusting the strategy according to the rewards of the external environment.

\textbf{Environment:} In this paper, environment primarily refers to the beamline. During interactions with this environment, the agent selects and executes an action based on the current state (output beam). Upon receiving an action, the environment transitions to a new state and provides a reward signal to the agent. The agent then uses this feedback to update its decision-making process, iteratively selecting subsequent actions until the maximum expected reward is achieved.

\textbf{State:} The state $ \mathbf{s} \in \mathbf{S}$ must contain sufficient information to capture changes at each step, enabling the agent to select the optimal action. In this study, the state is defined as the output beam of the beamline, which typically takes the shape of an ellipse. We denote the coordinates of its center position by $s_1$ and $s_2$, and the lengths of the semi-axes by $s_3$ and $s_4$.
\begin{equation}
    \mathbf{s}=[s_1,s_2,s_3,s_4].
\end{equation}

\textbf{Policy:} The policy function, $\mu(\mathbf{s})$, maps states to action, guiding the agent in selecting the next action within the environment.

\textbf{Action:} Given a state $\mathbf{s}_t$, the agent selects an action $a_t$ from a continuous action space $\mathcal{A}$. In the action space, the $t^{th}$ action is defined as the change in position and angle of the optical device in the beamline. Assume there are N optical devices, each of which includes 6 parameters: its position $(x,y,z)$ and angle $(\alpha,\beta,\gamma)$ denoted as:

\begin{equation}
\mathbf{a}=\{a^{1}_1,a^{1}_2,...,a^{1}_6,...,a^{N}_6 \}, \mathbf{a} \in \mathbb{R}^{6 \times N}.
\end{equation}

\textbf{Reward:} The result for the $t^{th}$ action is evaluated as a reward $r_t$. The goal of the task is to make the current state $ \mathbf{s_t}=[s^1_t,s^2_t,s^3_t,s^4_t]$ as close as possible to the target state $\mathbf{s_e}=[s^1_e,s^2_e,s^3_e,s^4_e]$, so we set the reward as follows:
\begin{equation}
    r_t= -WMAE, WMAE=[MAE([s^1_t,s^2_t],[s^1_e,s^2_e])+ \beta MAE([s^3_t,s^4_t],[s^3_e,s^4_e])],
\label{eq:WMAE}
\end{equation}
where MAE is Mean Absolute Error. In the experiment, the adjustment range of the radius is generally tiny, so we added a weight factor $\beta=2$ to control the output of the reward function.

\textbf{Episode:} an episode is one round for beam alignment, that
consists of a series of state $\mathbf{s}_t$, action $\mathbf{a}_t$, reward $r_t$, denoted as:
\begin{equation}
    [\mathbf{s}_0,\mathbf{a}_0,r_0,\mathbf{s}_1,\mathbf{a}_1,r_1,...,\mathbf{s}_t,\mathbf{a}_t,r_t,...,\mathbf{s}_n,\mathbf{a}_n,r_n].
\end{equation}
The process runs from the initial step to the terminal step. 
After each episode, the outcome is recorded, and the scenario is reinitialized.

\textbf{Return:} It is deﬁned as cumulative discount reward.
At step $t$, the return is formulated as:
\begin{equation}
    G_t=r_{t} + \gamma r_{t+1} + \gamma^2 r_{t+2} + ... ,
\label{eq:return}
\end{equation}
where $ 0 < \gamma < 1$ is called discount factor.

The goal of reinforcement learning is to maximize cumulative rewards over the long term. However, as shown in Equation (\ref{eq:return}), both the rewards and episode outcomes are uncertain, resulting in numerous possible scenarios where returns are variable. Consequently, the objective shifts to maximizing the expected cumulative rewards, represented by the following value function. The value function for a given state $\mathbf{s}$ is:
\begin{equation}
    V_{\pi}(\mathbf{s})=\mathbb{E}_{\pi}(G_t|\mathbf{s}_0=\mathbf{s}).
\end{equation}
and this value function is also called the state value function.
For given $\mathbf{s}$, $V_{\pi}(\mathbf{s})$ indicates the expected value of return
when following the policy $\pi$ starting from state $\mathbf{s}$. Besides, there is another kind of value function called
state-action value function:
\begin{equation}
    Q_{\pi}(\mathbf{s},\mathbf{a})=\mathbb{E}_{\pi}(G_t|\mathbf{s}_0=\mathbf{s},\mathbf{a}_0=\mathbf{a}).
\end{equation}
This indicates the expected return value when action $\mathbf{a}$ is taken from state $\mathbf{s}$ under policy $\pi$. With guidance from the value function, agents can purposefully accumulate scores and enhance their performance.

\section{Action-Attentive Deep Reinforcement Learning}

\begin{figure}[t]
    \centering
    \includegraphics[width=\linewidth]{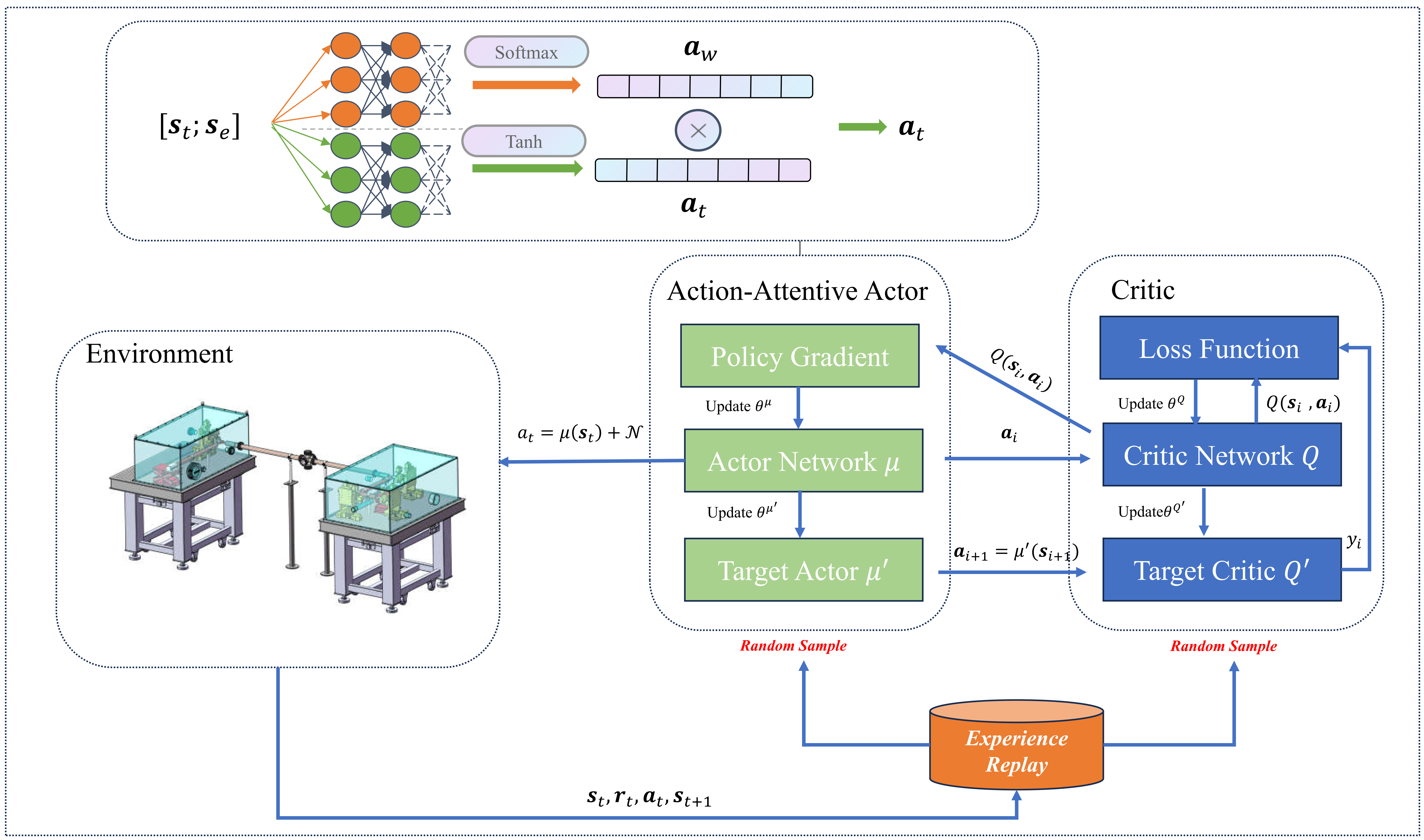}
    \caption{Our Approach for Autonomous Alignment of Beamlines.}
    \label{fig:enter-label}
\end{figure}

DeepMind \cite{mnih2013playing} initially combined deep neural networks with the Q-learning algorithm, introducing Deep Q-learning (DQN), a classic value-based reinforcement learning method. By leveraging the Bellman equation, DQN estimates Q-values for each action to derive the optimal policy, $\pi^*$. The Nature DQN \cite{mnih2015human} further improved stability and generalization by incorporating a target network and experience replay. However, DQN is primarily suitable for discrete action spaces, facing significant limitations in tasks requiring continuous control, such as robotic manipulation.

To address this limitation, researchers turned to policy gradient methods, directly optimizing policies to accommodate continuous action spaces. Building on this, Silver et al. \cite{silver2014deterministic} proposed the Deterministic Policy Gradient (DPG) algorithm, introducing the actor-critic framework for continuous action space problems. DPG uses a deterministic policy to generate actions, while a critic network evaluates action values. Later, Lillicrap et al. \cite{lillicrap2015continuous} combined DPG with deep neural networks, resulting in the Deep Deterministic Policy Gradient (DDPG) algorithm. By employing target networks, DDPG enhances training stability and demonstrates strong performance in complex continuous control tasks.

\subsection{Actor-Critic Structure}

Our method  is based on DDPG algorithm. The value function network, referred to as the critic network, takes the action and state as inputs $(\mathbf{s}, \mathbf{a})$ and outputs the Q-value $Q(\mathbf{s}, \mathbf{a})$. Additionally, another neural network, known as the actor network, approximates the policy function, with the state $\mathbf{s}$ as an input and action $\mathbf{a}$ as an output. Furthermore, target networks are utilized in the learning process to ensure parameter convergence.

Suppose the critic network is $Q(\mathbf{s},\mathbf{a}|\theta^{Q})$, its corresponding target critic network is  $Q'(\mathbf{s},\mathbf{a}|\theta^{Q'})$. the actor network is $\mu(\mathbf{s}|\theta^{\mu})$,
its corresponding target actor network is $\mu'(\mathbf{s}|\theta^{\mu'})$. $\theta^{\mu}$ and $\theta^{Q}$
are the weights for critic and actor networks, $\theta^{\mu'}$ and $\theta^{Q'}$ are target network weights.

\subsubsection{Actor Network}

Our method is off-policy, meaning that the policy used to generate a behavior (i.e., the policy that selects actions during training) and the policy used to evaluate the agent’s performance (i.e., the target policy) are not the same. Specifically, the action \( \mathbf{a}_{t} \) taken by the agent is not generated directly by the deterministic policy \( \mu(\mathbf{s}_t| \theta^\mu) \). To ensure sufficient exploration of the environment, we introduce exploration noise \( \mathcal{N} \) \cite{ladosz2022exploration} to the action selection process. This noise is added to the action as follows:

\begin{equation}
    a_t = \mu(\mathbf{s}_t|\theta^{\mu}) + \mathcal{N}.
\end{equation}

In each experiment, the desired output from the beamline may vary, requiring the agent to complete a series of similar yet distinct tasks. Traditional reinforcement learning algorithms can only identify a single target with one strategy, necessitating the training of multiple strategies for different target. To address this limitation, we introduce goal-oriented reinforcement learning (GoRL) \cite{pateria2021hierarchical}. Specifically, we incorporate the target state $\mathbf{s}_e$ into the policy function:

\begin{equation}
\begin{split}
  \mathbf{a}_t & = \mu([\mathbf{s}_t;\mathbf{s}_e]|\theta^{\mu}) + \mathcal{N}, \\
  \mu ([\mathbf{s}_t;&\mathbf{s}_e]|\theta^{\mu}) =MLP_1([\mathbf{s}_t;\mathbf{s}_e])
\end{split}
\label{eq:actor}
\end{equation}

\subsubsection{Critic Network}

The critic network and its corresponding target network share the same structure, multi-layer neural networks. We concatenate \(\mathbf{s}_t\) and \(\mathbf{a}_t\), and input them into a multi-layer perceptron (MLP) to generate the output, which is defined as:
\begin{equation}
    Q(\mathbf{s}_t,\mathbf{a}_t|\theta^{Q})=MLP_2([\mathbf{s}_t;\mathbf{a}_t]).
\end{equation}


\subsubsection{Hindsight Experience Replay} Additionally, experience replay collects and stores each agent's information in a memory pool for subsequent training of the actor and critic. Moreover, rewards in goal-oriented reinforcement learning are often sparse, as agents typically receive rewards only upon completing the goal, which is challenging in the early stages of training. To address this issue, we introduce hindsight experience replay (HER) \cite{andrychowicz2017hindsight,ren2019exploration,lin1992self,schaul2015prioritized} during training. HER significantly improves sample efficiency and accelerates learning, particularly in tasks such as robotic manipulation or navigation, where successful outcomes are rare.

\subsubsection{Updating Actor Network}

DDPG uses a deterministic policy \(  \mu(\mathbf{s}|\theta^\mu) \), which directly outputs a deterministic action \( \mathbf{a} = \mu(\mathbf{s}|\theta^\mu) \) for a given state \( \mathbf{s} \), without needing to sample from an action distribution \cite{silver2014deterministic}. Sampling $N$ tuples from the experience replay pool $\{(\mathbf{s}_i,\mathbf{a}_i,r_i,\mathbf{s}_{i+1})\}^N_{i=1}$, the goal is to optimize \( \theta^\mu \) to maximize the expected cumulative reward:
  \begin{equation}
       J(\theta^\mu) = \frac{1}{N} \sum_{i}^{N} \left[ Q(\mathbf{s}, \mu( \mathbf{s}|\theta^\mu)) |_{\mathbf{s}=\mathbf{s}_i}  \right].
  \end{equation}
The actor network parameters are updated through the policy gradient:

       \begin{equation}
       \theta^\mu \leftarrow \theta^\mu + \alpha_\mu \frac{1}{N} \sum_i^N \left[ \nabla_a Q(\mathbf{s}, \mathbf{a}|\theta^Q) |_{\mathbf{s}=\mathbf{s}_i,\mathbf{a}=\mu(\mathbf{s}_i)} \nabla_{\theta^\mu} \mu(\mathbf{s}|\theta^\mu) |_{\mathbf{s}=\mathbf{s}_i} \right].
       \end{equation}
The learning rate \( \alpha_\mu \) dictates how quickly the actor network’s parameters \( \theta^\mu \) are adjusted based on feedback from the critic. The actor updates its parameters to maximize the expected Q-value. 

\subsubsection{Updating Critic Network}
The critic network aims to minimize the mean squared error loss for the Q-value, where the target \( y_i \) is defined as:
\begin{equation}
      y_i = r_i + \gamma Q'(\mathbf{s}_{i+1}, \mu'(\mathbf{s}_{i+1}|\theta^{\mu'})|\theta^{Q'}),
\end{equation}
where \( r \) is the immediate reward and \( \gamma \) is the discount factor. The loss function for the critic network is:
  \begin{equation}
        L(\theta^Q) = \frac{1}{N} \sum_i^N \left[ \left( Q(\mathbf{s}_i, \mathbf{a}_i|\theta^Q) - y_i \right)^2 \right].
  \end{equation}
And, critic network parameters \( \theta^Q \) are updated as follows:
\begin{equation}
    \theta^Q \leftarrow \theta^Q - \alpha_Q \nabla_{\theta^Q} L(\theta^Q),
\end{equation}
where $\alpha_Q$ is learning rate.


\subsubsection{Updating Target Networks}

Furthermore, the target networks are updated at each step using a soft update method, which applies a small update. The following equations illustrate the updating process for the target networks:

\begin{equation}
\begin{aligned}
& \theta^{Q^{\prime}} \leftarrow \tau \theta^Q+(1-\tau) \theta^{Q^{\prime}}, \\
& \theta^{\mu^{\prime}} \leftarrow \tau \theta^\mu+(1-\tau) \theta^{\mu^{\prime}},
\end{aligned}
\end{equation}
where $\tau$ is the update parameter \cite{lillicrap2015continuous}.

\subsection{Action-Attentive Actor}

When the current state of the beamline approaches the target state, only minor adjustments are necessary. Conversely, significant modifications are required when the current state deviates considerably from the target. Furthermore, the optical elements must be prioritized during adjustments to the spot size and position differ entirely. This necessitates a policy function that can adjust the focus and amplitude for each step according to the specific task objectives. 

As a result, we redesign the actor network by deriving a hidden state that concatenates both the current and target states.
\begin{equation}
    \mathbf{h}_t= Relu(\mathbf{W}_2(\mathbf{W}_1 [\mathbf{s}_t;\mathbf{s}_e] + \textbf{b}_1)+\textbf{b}_2),
\end{equation}
where $\mathbf{s}_t$ is current state and $\mathbf{s}_e$ is target state. Inspired by \cite{vaswani2017attention}, we calculate the attention weight vector of an action based on $\mathbf{h}_t$:
\begin{equation}
    \mathbf{a}_{w}=Softmax(\mathbf{W}_3 \mathbf{h}_t+\mathbf{b}_3).
\end{equation}
Intuitively, the attention weights identify which optical components and their corresponding parameters need adjustment to transition the beamline from the current state to the target state. These attention weights are then applied to the output to generate the final action vector. Therefore, we rewrite Equation (\ref{eq:actor}) as:
\begin{equation}
    \mathbf{a}_t = \mathbf{a}_w Tanh(\mathbf{W}_4 \mathbf{h}_t+\mathbf{b}_4).
\end{equation}
\section{Experiments Setup}

\begin{figure}[t] 
  \begin{subfigure}[b]{0.5\linewidth}
    \centering
    \includegraphics[width=0.8\linewidth,height=7cm]{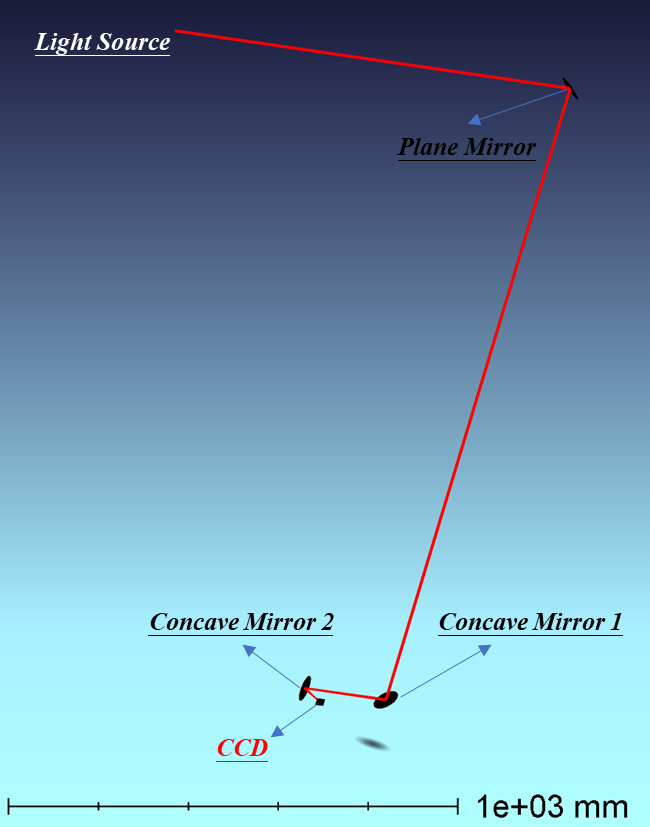} 
    \caption{System 1} 
    \label{fig:system1} 
  \end{subfigure}
  \begin{subfigure}[b]{0.5\linewidth}
    \centering
    \includegraphics[width=0.8\linewidth,height=7cm]{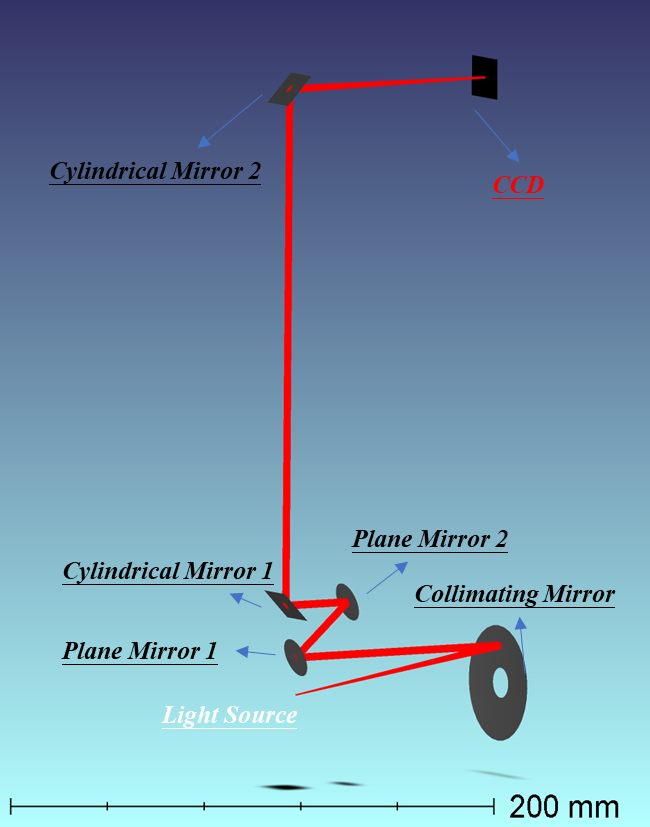} 
    \caption{System 2} 
    \label{fig:system2} 
  \end{subfigure} 
  \caption{Beamlines Structure.}
  \label{fig:experiment_system} 
\end{figure}

\subsection{Simulation Beamlines Construction}
Due to the high cost of real beamline equipment, we employ the simulation software Zemax\footnote{https://www.ansys.com/products/optics} to design two simulation beamlines to evaluate our proposed method.

The first system consists of one plane mirror, two concave mirrors, and a detector, as illustrated in Figure \ref{fig:experiment_system}. Each mirror has 6 adjustable parameters, while the output beam encompasses four parameters related to its position and size. Due to the long distance (1500 mm) from the plane mirror to concave mirror 1, the effective aperture of the optical element is relatively small (diameter 25.4 mm). Consequently, even a slight adjustment of the plane mirror may cause the laser beam to exceed the effective aperture, preventing it from being detected. To collect more valid data, the spatial position of the plane mirror is fixed during the actual process. As a result, the input parameters total $2 \times 6$, while the output parameters are 4. 

The second system consists of one collimating mirror, two plane mirrors, two cylindrical mirrors, and a detector. In this system, there are a total of 30 input parameters $5 \times 6$, while the output parameters remain at 4.

Additionally, since Zemax does not support direct interaction with Python, we collected thousands of data samples in Zemax to train a multi-layer perceptron (MLP) model to simulate the beamlines. This trained neural network was then used as the environment in our method.

\subsection{Evaluation Metrics}

For each experiment, we first initialize the environment and obtain the current state $ \mathbf{s}_t = [s^1_t, s^2_t, s^3_t, s^4_t] $. Next, we define the target state $ \mathbf{s}_e = [s^1_e, s^2_e, s^3_e, s^4_e] $ and adjust the parameters for optical devices in the simulation environment using the algorithm for $k$ iterations until the current state approaches the target state. In the experiment, we use WMAE (Equation \ref{eq:WMAE}) to evaluate the error. When $WMAE \leq \epsilon$, the model is considered to have found the target state.

Finally, we repeat the above experiment $N$ times, conducting $M$ experiments to reach the target state. The \textbf{first evaluation metric} can be defined as:
\begin{equation}
    coverage = \frac{M}{N}.
\end{equation}

Additionally, We define the number of algorithm iterations $k$ as \textbf{the second metric}. A larger number of iterations $k$ leads to decreased performance of the method, as more iterations are needed to reach the target state.

\subsection{Baselines}
Three baseline categories are selected for comparative analysis: the \textbf{Swarm Intelligence} algorithm, the \textbf{Bayesian Optimization} algorithm, and the \textbf{Reinforcement Learning-based} method.

\begin{itemize}
    \item \textbf{Differential Evolution (DE)} \cite{storn1997differential}  is a stochastic optimization algorithm for global optimization. DE is particularly effective for continuous space optimization problems and is widely used in fields like engineering design, machine learning, and control systems due to its simplicity and efficiency.
    \item \textbf{Genetic Algorithm (GA)} \cite{holland1992adaptation_ga} is an optimization technique based on natural selection and genetics. GA is commonly used to solve complex optimization problems, especially those challenging for traditional methods, such as combinatorial and function optimization.
    \item \textbf{Particle Swarm Optimization (PSO)} \cite{eberhart1995new_pso} is an optimization algorithm based on swarm intelligence. PSO simulates the foraging behavior of bird flocks, finding optimal solutions through information sharing among individuals.
    \item \textbf{Bayesian Optimization (BO)} \cite{snoek2012practical_bso}  is a sequential modeling approach for global optimization, particularly suitable for expensive black-box functions that lack direct gradient or structural information. It guides the search by constructing a posterior probability model of the target function, typically using a Gaussian process (GP).
    \item \textbf{Deep Deterministic Policy Gradient (DDPG) } \cite{lillicrap2015continuous} is a reinforcement learning algorithm for solving continuous action space problems. DDPG combines deep learning with policy gradient methods and can handle tasks with high-dimensional state and action spaces.
\end{itemize}

Additionally, we constructed a variant model in which the actor network did not incorporate action-attentive mechanism, namely treating each component of the action vector equally. In this actor network , the action $\mathbf{a}_t$ is computed by:
\begin{equation}
    \mathbf{a}_t =\frac{1}{N} Tanh(\mathbf{W}_4 \mathbf{h}_t+\mathbf{b}_4),
\end{equation}
where $N$ denotes the dimension of $\mathbf{h}_t$.

\begin{table}[t]\scriptsize 
\centering
\caption{Baselines Comparison. * represents the baseline of our implementation by bayesianoptimization toolkit \cite{github_bso} and scikit-opt \cite{github_opt}. We highlight the best performance among all methods in bold. $cov$ represents the metric coverage. \textbf{\textit{ w/o att}} denotes our variant model. $max(k)=10$ means that the algorithm executes a maximum of 10 iterations, and $avg(k)$ defines how many iterations are needed on average to find the target state. If the target state is not found in 10 iterations, k=10.}
\begin{tabular}{ccccccccccccc}
\hline
\multirow{4}{*}{\textbf{Models}} &
  \multicolumn{12}{c}{\textbf{System1}} \\ \cline{2-13} 
 &
  \multicolumn{6}{c}{$\mathbf{\epsilon=0.05}$} &
  \multicolumn{6}{c}{$\mathbf{\epsilon=0.1}$} \\\cline{2-13} 
 &
  \multicolumn{2}{c}{$\mathbf{max(k)=10}$}     &
  \multicolumn{2}{c}{$\mathbf{max(k)=20}$}    &
  \multicolumn{2}{c}{$\mathbf{max(k)=50}$} &
  \multicolumn{2}{c}{$\mathbf{max(k)=10}$} &
  \multicolumn{2}{c}{$\mathbf{max(k)=20}$} &
  \multicolumn{2}{c}{$\mathbf{max(k)=50}$} \\ \cline{2-13} 
 &
  $cov$ &
  $avg(k)$ &
  $cov$ &
  $avg(k)$ &
  $cov$ &
  $avg(k)$ &
  $cov$ &
  $avg(k)$ &
  $cov$ &
  $avg(k)$ &
  $cov$ &
  $avg(k)$ \\ \hline
DE* &
  0.014 &
  9.973 &
  0.101 &
  19.380 &
  0.317 &
  42.685 &
  0.223 &
  9.309 &
  0.562 &
  15.285 &
  0.830 &
  23.406 \\
GA* &
  0.085 &
  9.784 &
  0.214 &
  18.344 &
  0.307 &
  40.238 &
  0.417 &
  8.607 &
  0.683 &
  12.731 &
  0.794 &
  19.732 \\
PSO* &
  0.029 &
  9.952 &
  0.180 &
  18.948 &
  0.329 &
  40.151 &
  0.263 &
  9.386 &
  0.646 &
  14.520 &
  0.752 &
  22.827 \\
BSO* &
  0.001 &
  9.993 &
  0.001 &
  19.990 &
  0.010 &
  49.779 &
  0.015 &
  9.945 &
  0.033 &
  19.755 &
  0.117 &
  47.455 \\
DDPG &
  0.445 &
  7.721 &
  0.518 &
  12.825 &
  0.557 &
  26.601 &
  0.924 &
  3.564 &
  0.954 &
  4.131 &
  0.961 &
  5.388 \\
OURS &
  \textbf{0.744} &
  \textbf{5.908} &
  \textbf{0.899} &
  \textbf{7.463} &
  \textbf{0.944} &
  \textbf{9.675} &
  \textbf{0.956} &
  \textbf{3.063} &
  \textbf{0.993} &
  \textbf{3.248} &
  \textbf{0.999} &
  \textbf{3.308} \\
\textit{-w/o att} &
  0.380 &
  8.520 &
  0.615 &
  13.565 &
  0.855 &
  20.319 &
  0.746 &
  6.463 &
  0.885 &
  8.197 &
  0.983 &
  9.667 \\ \hline
\multirow{4}{*}{\textbf{Models}} &
  \multicolumn{12}{c}{\textbf{System2}} \\ \cline{2-13} 
 &
  \multicolumn{6}{c}{$\mathbf{\epsilon=0.05}$} &
  \multicolumn{6}{c}{$\mathbf{\epsilon=0.1}$} \\\cline{2-13} 
 &
  \multicolumn{2}{c}{$\mathbf{max(k)=10}$}     &
  \multicolumn{2}{c}{$\mathbf{max(k)=20}$}    &
  \multicolumn{2}{c}{$\mathbf{max(k)=50}$} &
  \multicolumn{2}{c}{$\mathbf{max(k)=10}$} &
  \multicolumn{2}{c}{$\mathbf{max(k)=20}$} &
  \multicolumn{2}{c}{$\mathbf{max(k)=50}$} \\ \cline{2-13}
 &
 $cov$ &
  $avg(k)$ &
  $cov$ &
  $avg(k)$ &
  $cov$ &
  $avg(k)$ &
  $cov$ &
  $avg(k)$ &
  $cov$ &
  $avg(k)$ &
  $cov$ &
  $avg(k)$ \\ \hline
DE* &
  0.089 &
  9.697 &
  0.241 &
  18.013 &
  0.403 &
  38.325 &
  0.396 &
  8.141 &
  0.589 &
  13.320 &
  0.819 &
  21.389 \\
GA* &
  0.246 &
  9.224 &
  0.372 &
  15.947 &
  0.479 &
  33.053 &
  0.572 &
  7.279 &
  0.744 &
  10.456 &
  0.881 &
  15.433 \\
PSO* &
  0.265 &
  9.298 &
  0.448 &
  15.496 &
  0.507 &
  31.279 &
  0.658 &
  7.269 &
  0.809 &
  9.874 &
  0.897 &
  13.272 \\
BSO* &
  0.009 &
  9.979 &
  0.028 &
  19.777 &
  0.083 &
  47.951 &
  0.099 &
  9.595 &
  0.172 &
  18.330 &
  0.371 &
  39.070
   \\
DDPG &
  0.084 &
  9.533 &
  0.113 &
  18.533 &
  0.144 &
  44.551 &
  0.509 &
  6.811 &
  0.567 &
  11.381 &
  0.621 &
  23.513 \\
OURS &
  \textbf{0.804} &
  \textbf{5.631} &
  \textbf{0.895} &
  \textbf{7.029} &
  \textbf{0.928} &
  \textbf{9.477} &
  \textbf{0.965} &
  \textbf{3.002} &
  \textbf{0.981} &
  \textbf{3.248} &
  0.985 &
  \textbf{3.743} \\
\textit{-w/o att} &
  0.239 &
  9.166 &
  0.533 &
  15.421 &
  0.855 &
  22.845 &
  0.662 &
  7.485 &
  0.913 &
  9.499 &
  \textbf{0.997} &
  10.245 \\ \hline
\end{tabular}
\label{tab:baselines}
\end{table}

\section{Results and Analysis}

\subsection{Baselines Comparison}

For each method and setting, we conduct 500 random experiments, starting with an initial state of the environment and subsequently adjusting the parameters to reach a random target state. To mitigate the effects of randomness, we employ different seeds and repeat the experiments three times, calculating the average results. The outcomes are presented in Table \ref{tab:baselines}.  

The Table \ref{tab:baselines} indicates that the swarm evolution algorithms can yield favorable results with a higher number of iterations when the threshold $\epsilon$ is high. For instance, in System 1, the genetic algorithm (GA) achieves a coverage rate of $0.794$ with an average of $19.732$ iterations, when $\epsilon=0.05,max(k)=50$. That is to say, when the threshold is high, most experiments can find the target after about 20 iterations. However, when the threshold is low ($\epsilon=0.05,max(k)=50$), in System 1, for example, the genetic algorithm (GA) only attains a coverage rate of $0.307$, requiring approximately $40$ iterations.

Bayesian optimization (BO) methods have achieved good results in many optimization fields. However, in our task, BO does not achieve good results, and its effect is worse than all the swarm evolution algorithms.

Since we adopted the off-policy reinforcement learning method, that is, we used historical data to train the model, and finally only performed inference in the experiment. Therefore, the reinforcement learning method is superior to other types of methods in terms of iteration steps and coverage. It can be seen that the reinforcement learning method only needs about 5 steps on average to find 480 ($cov=0.961$) target states in system 1, when $\epsilon=0.1,max(k)=50$.

Finally, our model demonstrates significant performance improvements in both systems compared to other methods, particularly in the average number of iterations, which decreased notably. For instance, in System 2 ($\epsilon=0.1, max(k)=50$) , the DDPG-based reinforcement learning method requires an average of 23 steps to achieve a coverage of 0.621, whereas our model reaches a coverage of 0.985 in just 3 steps. 

Through comparative experiments with the baseline, the following conclusions can be drawn: First, with sufficient iterations, the evolutionary algorithm demonstrates competitive performance on this task. Second, off-policy reinforcement learning significantly improves both speed and accuracy. Finally, our method outperforms all others, achieving the best performance on both simulation systems.

\begin{figure}[t] 
  \begin{subfigure}[b]{0.328\linewidth}
    \centering
    \includegraphics[width=\linewidth,height=2.5cm]{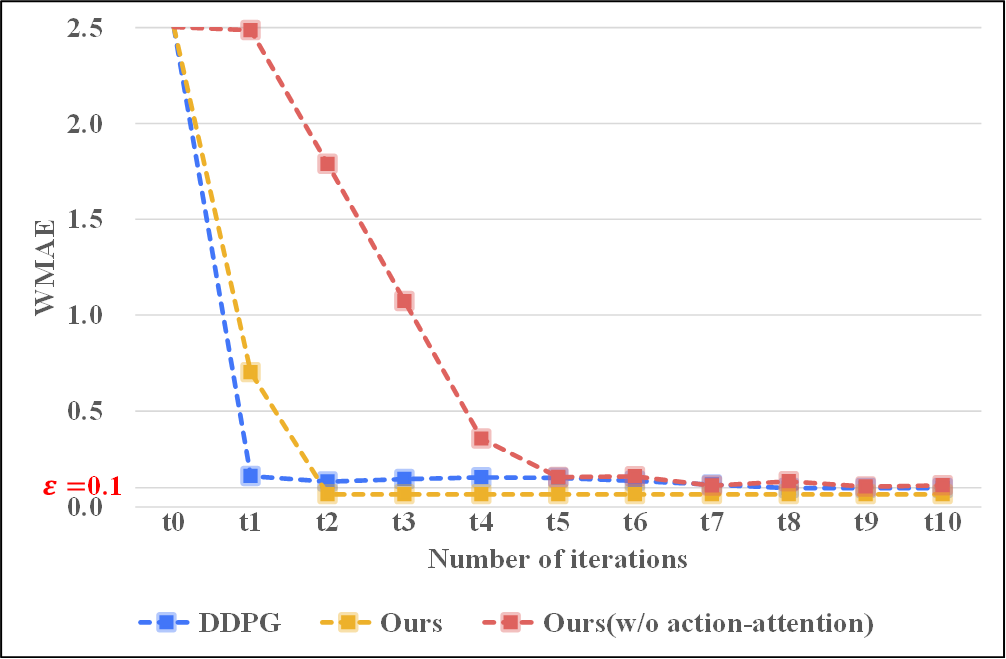} 
    \caption{Case 1} 
    \label{fig:iteration_case1} 
  \end{subfigure}
  \begin{subfigure}[b]{0.328\linewidth}
    \centering
    \includegraphics[width=\linewidth,height=2.5cm]{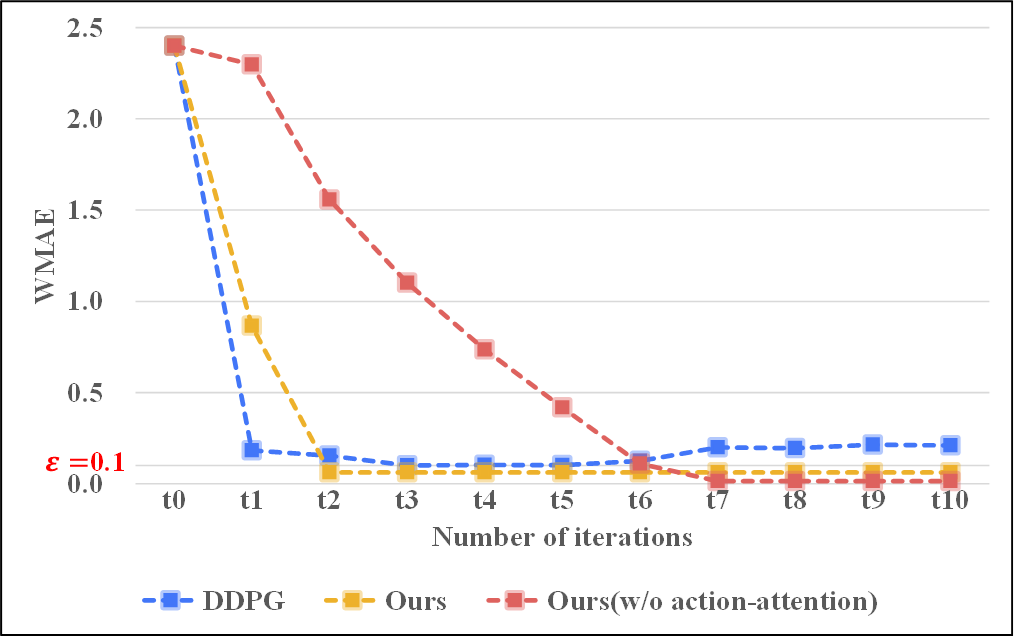} 
    \caption{Case 2} 
    \label{fig:iteration_case2} 
  \end{subfigure} 
  \begin{subfigure}[b]{0.328\linewidth}
    \centering
    \includegraphics[width=\linewidth,height=2.5cm]{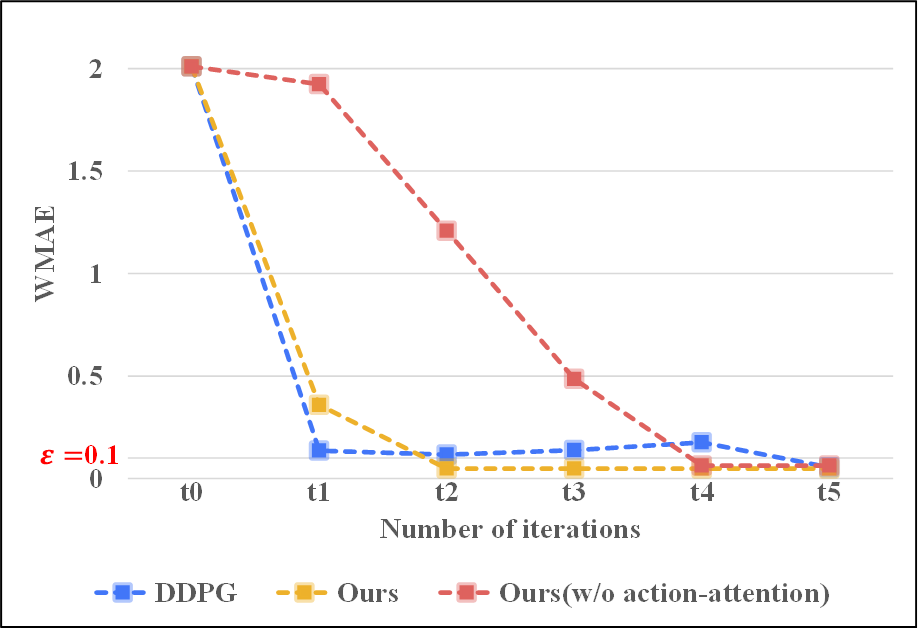} 
    \caption{Case 3} 
    \label{fig:iteration_case3} 
  \end{subfigure} 
  \caption{Case study, we use three algorithms starting from the same initial state and setting the same target state, with a maximum number of iterations of 10 and $\epsilon=0.1$.}
  \label{fig:iteration_case} 
\end{figure}

\subsection{Ablation study}

This section analyzes the proposed model to evaluate the contribution of action attentive mechanism. As shown in Table \ref{tab:baselines}, replacing the actor (which lacks the action-attention mechanism, denoted as \textit{w/o att}) resulted in a significant drop in model performance, requiring more iterations to reach the target state. Nevertheless, this modified model still outperforms the DDPG-based reinforcement learning algorithm in the baselines. The ablation experiment demonstrates the feasibility of the proposed motivation and the effectiveness of the action-attentive actor.

\subsection{Case Study} 

\subsubsection{Iteration Visualization}

We take system 2 as an example and select 3 steps of data. We calculate the WMAE (Computed by Equation \ref{eq:WMAE}) of the output state and the target state after each iteration, the results are shown in Figure \ref{fig:iteration_case}. It can be seen that the action attentive actor reaches the target state after 2 rounds of iterations, in case 1. 
Although the DDPG-based RL algorithm also reaches the target state after two rounds of iterations, the WMAE does not decrease after further iterations but increases slightly. 
Without the action-attentive actor, the target state can still be reached, but it requires a greater number of iterations.

Through iteration visualization, we can draw the following conclusions: the action-attentive actor can more accurately identify the direction and magnitude of action adjustments, enabling the model to reach the target state more rapidly.

\subsubsection{Attention Visualization}

\begin{figure}[t] 
  \begin{subfigure}[b]{0.328\linewidth}
    \centering
    \includegraphics[width=\linewidth]{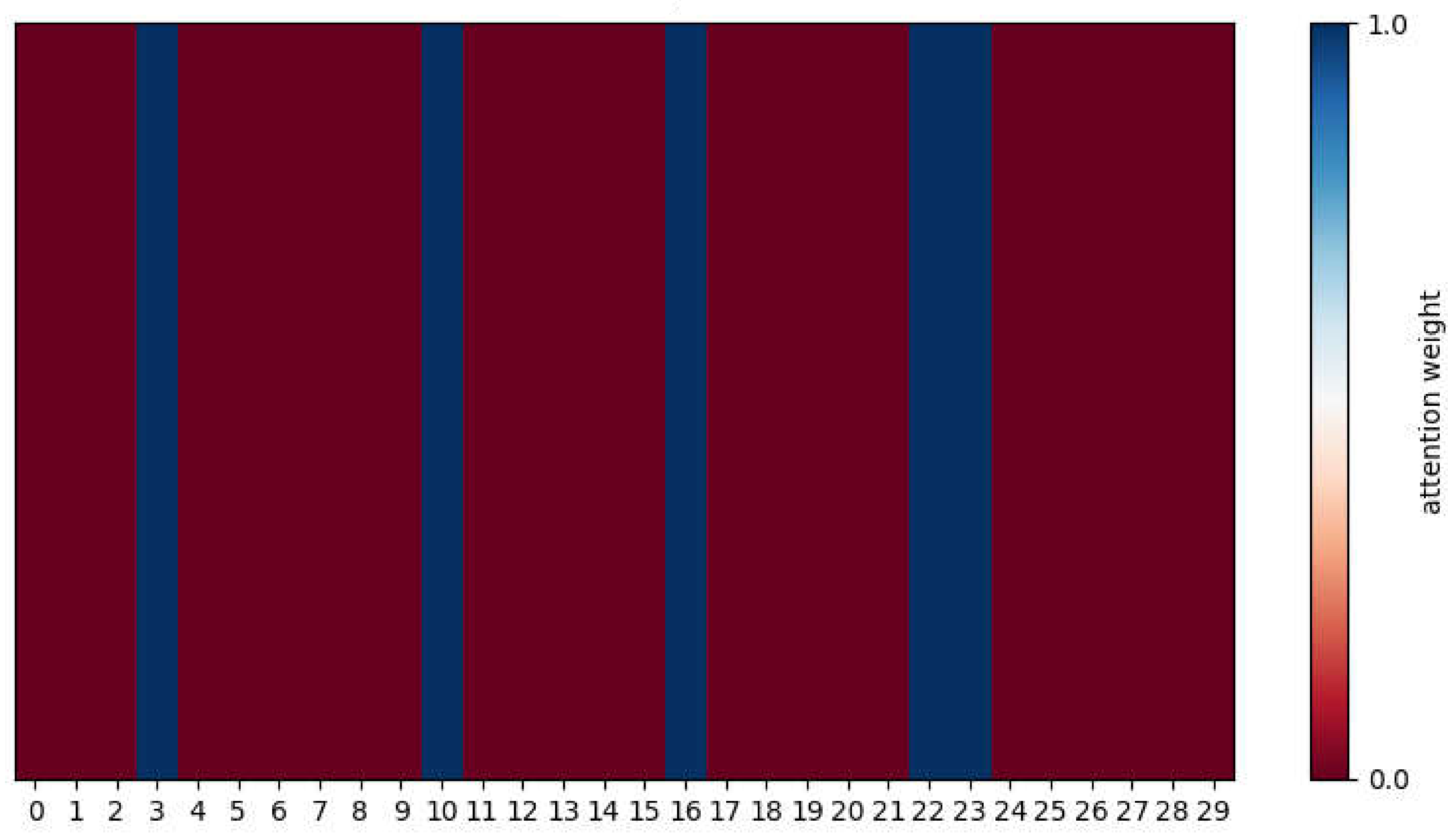} 
    \caption{Step 1} 
    \label{fig:attention_case1} 
  \end{subfigure}
  \begin{subfigure}[b]{0.328\linewidth}
    \centering
    \includegraphics[width=\linewidth]{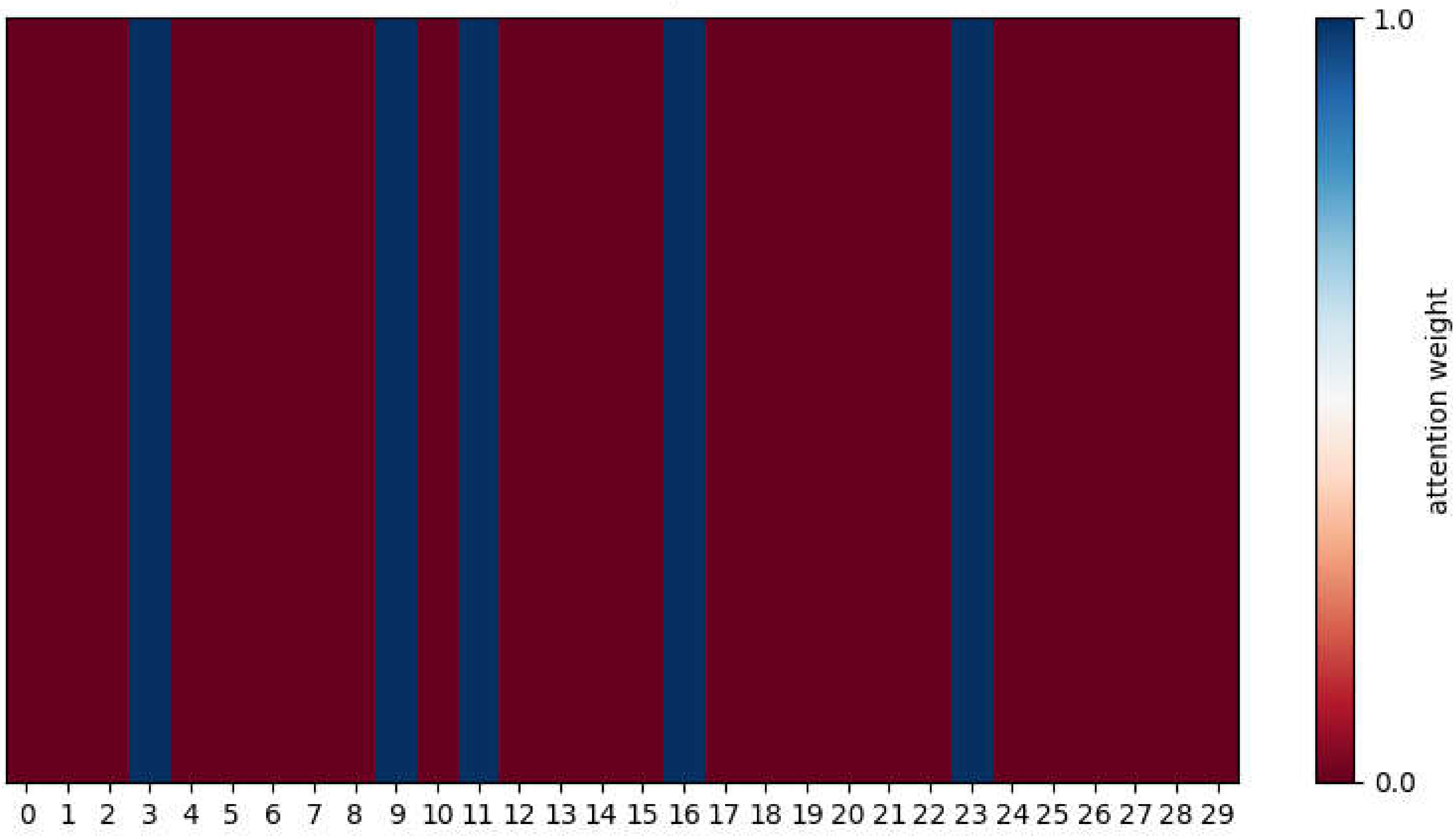} 
    \caption{Step 2} 
    \label{fig:attention_case2} 
  \end{subfigure} 
  \begin{subfigure}[b]{0.328\linewidth}
    \centering
    \includegraphics[width=\linewidth]{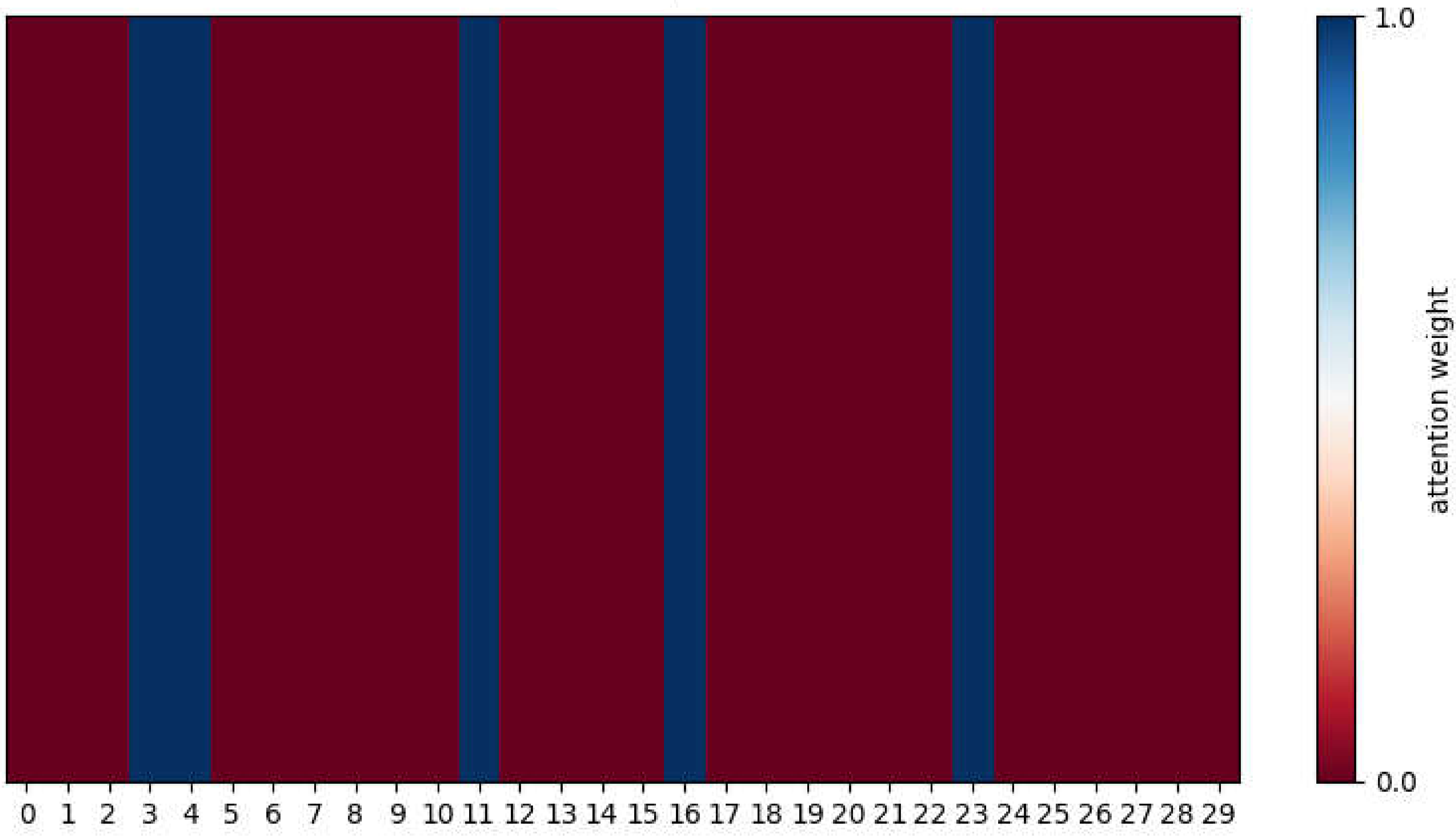} 
    \caption{Step 3} 
    \label{fig:attention_case3} 
  \end{subfigure} 
  \caption{Action-attention visualization. In this case, ours model reaches the target state through 3 steps from the initial state. $\{0-30\}$ represent the parameters of the optical devices in the beamline, for example, $\{0-5\}$ represents the position and angle of the first device. In the figure, the blue part indicates that the attention weight is greater than 0.01.}
  \label{fig:attention_case} 
\end{figure}

Generally, an experienced engineer adjusts a beamline to reach the target state through a continuous process. They typically begin by adjusting the position of the spot, followed by the spot size, and then proceed to fine-tuning. Consequently, the optical devices adjust at each step differ. We investigate whether the trained action-attentive actor network can produce similar strategies. We visualize the attention weights of the actor at each step, with the results presented in Figure \ref{fig:attention_case}. In the first iteration, the model's strategy focuses on adjusting parameters $\{2-4\}$, $\{9-11\}$, and $\{21-24\}$, while in the second iteration, it prioritizes different parameters. This observation indicates that through training, the actor can dynamically adapt its strategy based on the current state, enabling it to quickly find the target state.

\section{Conclusion}

This paper models autonomous  alignment of beamlines as a MDP, employs reinforcement learning, and develops an intelligent agent capable of optimizing the configuration of optical components. The key characteristics of beamline adjustments—sequential multi-step operations, varying degrees of adjustments based on output proximity to target states, and the distinct impacts of specific optical components on beam properties were effectively addressed through our approach. The introduction of a policy network based on action attention further enhances the agent's ability to generate precise adjustment actions, significantly improving both the efficiency and accuracy of the adjustment process. Our simulations demonstrated the method's effectiveness, paving the way for more automated and precise beamline operations in various scientific disciplines, including materials science, biology, and chemistry. Future work will focus on refining this approach and exploring its application to a broader range of experimental scenarios, ultimately contributing to the advancement of synchrotron radiation technology.

\bibliographystyle{splncs04}
\bibliography{ref}

\begin{thebibliography}{10}
\providecommand{\url}[1]{\texttt{#1}}
\providecommand{\urlprefix}{URL }
\providecommand{\doi}[1]{https://doi.org/#1}

\bibitem{andrychowicz2017hindsight}
Andrychowicz, M., Wolski, F., Ray, A., Schneider, J., Fong, R., Welinder, P., McGrew, B., Tobin, J., Pieter~Abbeel, O., Zaremba, W.: Hindsight experience replay. Advances in neural information processing systems  \textbf{30} (2017)

\bibitem{boltz2020feedback}
Boltz, T., Brosi, M., Br{\"u}ndermann, E., Haerer, B., Kaiser, P., Pohl, C., Schreiber, P., Yan, M., Asfour, T., M{\"u}ller, A.S.: Feedback design for control of the micro-bunching instability based on reinforcement learning. In: CERN Yellow Reports: Conference Proceedings. vol.~9, pp. 227--227 (2020)

\bibitem{chen2023trend}
Chen, X., Qi, X., Su, C., He, Y., Wang, Z., Sun, K., Jin, C., Chen, W., Liu, S., Zhao, X., et~al.: Trend-based sac beam control method with zero-shot in superconducting linear accelerator. arXiv preprint arXiv:2305.13869  (2023)

\bibitem{eberhart1995new_pso}
Eberhart, R., Kennedy, J.: A new optimizer using particle swarm theory. In: MHS'95. Proceedings of the sixth international symposium on micro machine and human science. pp. 39--43. Ieee (1995)

\bibitem{garcia2016synchrotron}
Garc{\'\i}a, G.: Synchrotron radiation: basics, methods and applications (2016)

\bibitem{github_opt}
Guo, F.: Swarm intelligence in python (2017--), \url{https://github.com/guofei9987/scikit-opt/}

\bibitem{holland1992adaptation_ga}
Holland, J.H.: Adaptation in natural and artificial systems: an introductory analysis with applications to biology, control, and artificial intelligence. MIT press (1992)

\bibitem{hwang2022beam}
Hwang, K., Maruta, T., Plastun, A., Fukushima, K., Zhang, T., Zhao, Q., Ostroumov, P., Nash, S.: Beam tuning at the frib front end using machine learning. Proc. IPAC  \textbf{22},  983--986 (2022)

\bibitem{kaiser2023learning}
Kaiser, J., Xu, C., Eichler, A., Garcia, A.S., Stein, O., Br{\"u}ndermann, E., Kuropka, W., Dinter, H., Mayet, F., Vinatier, T., et~al.: Learning to do or learning while doing: Reinforcement learning and bayesian optimisation for online continuous tuning. arXiv preprint arXiv:2306.03739  (2023)

\bibitem{karaca2024optimization}
Karaca, A.S., Bostanci, E., Ketenoglu, D., Harder, M., Canbay, A.C., Ketenoglu, B., Eren, E., Aydin, A., Yin, Z., Guzel, M.S., et~al.: Optimization of synchrotron radiation parameters using swarm intelligence and evolutionary algorithms. Journal of Synchrotron Radiation  \textbf{31}(2) (2024)

\bibitem{ladosz2022exploration}
Ladosz, P., Weng, L., Kim, M., Oh, H.: Exploration in deep reinforcement learning: A survey. Information Fusion  \textbf{85},  1--22 (2022)

\bibitem{lillicrap2015continuous}
Lillicrap, T.: Continuous control with deep reinforcement learning. arXiv preprint arXiv:1509.02971  (2015)

\bibitem{lin1992self}
Lin, L.J.: Self-improving reactive agents based on reinforcement learning, planning and teaching. Machine learning  \textbf{8},  293--321 (1992)

\bibitem{mazyavkina2021reinforcement}
Mazyavkina, N., Sviridov, S., Ivanov, S., Burnaev, E.: Reinforcement learning for combinatorial optimization: A survey. Computers \& Operations Research  \textbf{134},  105400 (2021)

\bibitem{meier2012orbit}
Meier, E., Tan, Y., LeBlanc, G., et~al.: Orbit correction studies using neural networks. In: Proc. 3rd Int. Particle Accelerator Conf.(IPAC’12). pp. 2837--2839 (2012)

\bibitem{mnih2013playing}
Mnih, V.: Playing atari with deep reinforcement learning. arXiv preprint arXiv:1312.5602  (2013)

\bibitem{mnih2015human}
Mnih, V., Kavukcuoglu, K., Silver, D., Rusu, A.A., Veness, J., Bellemare, M.G., Graves, A., Riedmiller, M., Fidjeland, A.K., Ostrovski, G., et~al.: Human-level control through deep reinforcement learning. nature  \textbf{518}(7540),  529--533 (2015)

\bibitem{morris2024general}
Morris, T., Rakitin, M., Islegen-Wojdyla, A., Du, Y., Fedurin, M., Giles, A., Leshchev, D., Li, W., Moeller, P., Nash, B., et~al.: A general bayesian algorithm for the autonomous alignment of beamlines. arXiv preprint arXiv:2402.16716  (2024)

\bibitem{morris2022fly}
Morris, T., Rakitin, M., Giles, A., Lynch, J., Walter, A.L., Nash, B., Abell, D., Moeller, P., Pogorelov, I., Goldring, N.: On-the-fly optimization of synchrotron beamlines using machine learning. In: Optical System Alignment, Tolerancing, and Verification XIV. vol. 12222, pp. 171--175. SPIE (2022)

\bibitem{github_bso}
Nogueira, F.: {Bayesian Optimization}: Open source constrained global optimization tool for {Python} (2014--), \url{https://github.com/bayesian-optimization/BayesianOptimization}

\bibitem{pateria2021hierarchical}
Pateria, S., Subagdja, B., Tan, A.h., Quek, C.: Hierarchical reinforcement learning: A comprehensive survey. ACM Computing Surveys (CSUR)  \textbf{54}(5),  1--35 (2021)

\bibitem{puterman2014markov}
Puterman, M.L.: Markov decision processes: discrete stochastic dynamic programming. John Wiley \& Sons (2014)

\bibitem{ren2019exploration}
Ren, Z., Dong, K., Zhou, Y., Liu, Q., Peng, J.: Exploration via hindsight goal generation. Advances in Neural Information Processing Systems  \textbf{32} (2019)

\bibitem{ruichun2021application}
Ruichun, L., Qinglei, Z., Qingru, M., Bocheng, J., Kun, W., Changliang, L., Zhentang, Z.: Application of machine learning in orbital correction of storage ring. High Power Laser and Particle Beams  \textbf{33}(3),  034007--1 (2021)

\bibitem{schaul2015prioritized}
Schaul, T.: Prioritized experience replay. arXiv preprint arXiv:1511.05952  (2015)

\bibitem{schirmer2019orbit}
Schirmer, D., et~al.: Orbit correction with machine learning techniques at the synchrotron light source delta. In: Proc. of 17th Int. Conf. on Accelerator and Large Experimental Physics Control Systems (ICALEPCS’19). pp. 1426--1430 (2019)

\bibitem{silver2014deterministic}
Silver, D., Lever, G., Heess, N., Degris, T., Wierstra, D., Riedmiller, M.: Deterministic policy gradient algorithms. In: International conference on machine learning. pp. 387--395. Pmlr (2014)

\bibitem{snoek2012practical_bso}
Snoek, J., Larochelle, H., Adams, R.P.: Practical bayesian optimization of machine learning algorithms. Advances in neural information processing systems  \textbf{25} (2012)

\bibitem{storn1997differential}
Storn, R., Price, K.: Differential evolution--a simple and efficient heuristic for global optimization over continuous spaces. Journal of global optimization  \textbf{11},  341--359 (1997)

\bibitem{vaswani2017attention}
Vaswani, A.: Attention is all you need. Advances in Neural Information Processing Systems  (2017)

\bibitem{zhang2023multi}
Zhang, J., Qi, P., Wang, J.: Multi-objective genetic algorithm for synchrotron radiation beamline optimization. Journal of Synchrotron Radiation  \textbf{30}(1),  51--56 (2023)

\end{thebibliography}

\end{document}